\documentclass[sigconf]{acmart}

\usepackage{balance}

\usepackage{booktabs} %
\usepackage{cleveref}
\usepackage{url}

\graphicspath{{./figures/}}
\usepackage{graphicx}
\usepackage{subcaption}

\usepackage{xspace}
\newcommand*{\eg}{e.g.,\@\xspace}
\newcommand*{\ie}{i.e.,\@\xspace}

\usepackage{xcolor}

\newcommand{\sparagraph}[1]{\vspace{1mm}\noindent {\bf #1}}

\AtBeginDocument{%
  \providecommand\BibTeX{{%
    \normalfont B\kern-0.5em{\scshape i\kern-0.25em b}\kern-0.8em\TeX}}}

\copyrightyear{2021} 
\acmYear{2021} 
\setcopyright{acmlicensed}\acmConference[aiDM'21]{Fourth Workshop in Exploiting AI Techniques for Data Management }{June 25, 2021}{Xi'an, Shaanxi, China}
\acmBooktitle{Fourth Workshop in Exploiting AI Techniques for Data Management (aiDM'21), June 25, 2021, Xi'an, Shaanxi, China}
\acmPrice{15.00}
\acmDOI{} %
\acmISBN{} %

\setcopyright{rightsretained}

\begin{document}

\copyrightyear{2021}
\acmYear{2021}
\acmConference[aiDM'21]{Fourth Workshop in Exploiting AI Techniques for Data Management }{June 20--25, 2021}{Virtual Event, China}
\acmBooktitle{Fourth Workshop in Exploiting AI Techniques for Data Management (aiDM'21), June 20--25, 2021, Virtual Event, China}\acmDOI{10.1145/3464509.3464885}
\acmISBN{978-1-4503-8535-0/21/06}

\title{LEA: A Learned Encoding Advisor for Column Stores}

\settopmatter{authorsperrow=4}

\author{Lujing Cen}
\affiliation{%
    \institution{MIT CSAIL}
    \city{Cambridge}
    \state{MA}
    \country{USA}
}
\email{lujing@mit.edu}

\author{Andreas Kipf}
\affiliation{%
    \institution{MIT CSAIL}
    \city{Cambridge}
    \state{MA}
    \country{USA}
}
\email{kipf@mit.edu}

\author{Ryan Marcus}
\affiliation{%
    \institution{MIT CSAIL, Intel Labs}
    \city{Cambridge}
    \state{MA}
    \country{USA}
}
\email{ryanmarcus@mit.edu}

\author{Tim Kraska}
\affiliation{%
    \institution{MIT CSAIL}
    \city{Cambridge}
    \state{MA}
    \country{USA}
}
\email{kraska@mit.edu}

\renewcommand{\shortauthors}{Cen et al.}

\begin{abstract}
Data warehouses organize data in a columnar format to enable faster scans and better compression. Modern systems offer a variety of column encodings that can reduce storage footprint and improve query performance. Selecting a good encoding scheme for a particular column is an optimization problem that depends on the data, the query workload, and the underlying hardware.

We introduce Learned Encoding Advisor (LEA), a learned approach to column encoding selection.
LEA is trained on synthetic datasets with various distributions on the target system.
Once trained, LEA uses sample data and statistics (such as cardinality) from the user's database  to predict the optimal column encodings.
LEA can optimize for encoded size, query performance, or a combination of the two.
Compared to the heuristic-based encoding advisor of a commercial column store on TPC-H, LEA achieves 19\% lower query latency while using 26\% less space.
\end{abstract}

\maketitle

\section{Introduction}
Column stores offer different encodings, such as delta or dictionary, as well as general-purpose compression schemes (\eg ZSTD~\cite{zstd}). The user specifies the encoding for each column or allows the database to apply a set of default encodings based on the column type. Notably, a few systems use heuristic-based approaches to select column encodings. For example, Amazon Redshift~\cite{redshift-encodings} and Vertica~\cite{vertica-encodings} select encodings by sampling contiguous rows from a table. SingleStore supports automatic encoding selection for each column segment, but it is not clear what heuristic is used~\cite{singlestore-encodings}.

In this paper, we introduce the Learned Encoding Advisor (LEA).
LEA addresses three limitations of the aforementioned approaches:
First, LEA not only optimizes for encoded size, but also considers end-to-end scan performance on the target system, including I/O and in-memory decompression.
Second, LEA does \emph{not} constrain its optimization to finding a single encoding that works for the entire column.
Instead, it allows for an encoding per column \emph{and} block, exploiting localized correlations in data.
Third, LEA considers both data statistics (\eg cardinality, min/max) and sample statistics during optimization.
We show that only using sample data to select the best encoding is insufficient for certain schemes. For example, dictionary and frame-of-reference (FOR) encodings depend on cardinality and domain size, respectively, neither of which can be adequately captured by sampling.
Finally, instead of requiring re-tuning for different CPUs and storage devices, LEA uses a two-part training process that allows it to quickly adapt itself to the user's underlying hardware.

We integrate LEA with a commercial column store and demonstrate that it improves upon the built-in encoding advisor on two workloads, even when a single encoding strategy is used for each column.
Specifically, LEA achieves 19\% lower cold-cache query latency on TPC-H while using 26\% less space.
We also compare LEA against the optimal set of column encodings (brute-forced) and show that LEA always stays within 10\% of the optimal encoding, regardless of whether we optimize for size or query latency.
In its current form, LEA assumes a uniform workload during training and inference.
In future work, we plan to consider the concrete access patterns of a given workload to unlock its full potential.

\sparagraph{Related Work.}
There is limited work in machine learning techniques for choosing optimal encodings. A project known as shrynk feeds statistics about a data frame into a classification model to predict the best compression scheme~\cite{shrynk}. The work, while specific to pandas~\cite{pandas}, demonstrates that it is possible to apply machine learning in the context of encoding selection. However, shrynk requires the optimization objective to be defined during training and only supports selection on the granularity of an entire file.

Other approaches to improving database performance include automated tuning, where different configuration parameters exposed by the database are adjusted using heuristics or reinforcement learning~\cite{tuning1, tuning2}. There is also extensive work in column advisors that automatically determine the best partitions, column clusters, indexes, and materialized views~\cite{DBLP:conf/edbt/RasinZ13, zilio}. \cite{DBLP:conf/edbt/RasinZ13} tries to select the optimal column encodings, but does so using fixed heuristics based on the cardinality of the data and whether or not a column is sorted. Additionally, a few systems perform operations directly on encoded data by pushing down predicates and aggregates~\cite{memsql-encoding-selection, powerdrill}.

\begin{figure*}[ht]
\includegraphics[width=0.9\linewidth]{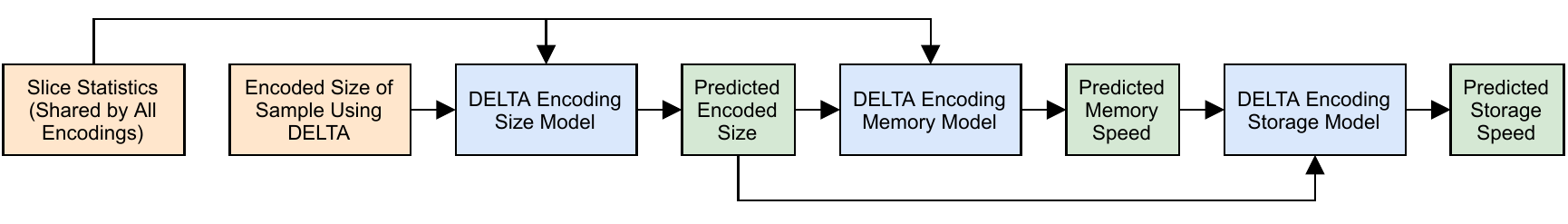}
\centering
\caption{Predicting encoded size, in-memory scan speed, and from-storage scan speed for a single column slice. The size model consumes slice statistics and the encoded sample size. The in-memory scan speed model consumes the predicted size and slice statistics. The from-storage scan speed model consumes the predicted size and the predicted in-memory scan speed.}
\label{fig:diagram}
\end{figure*}

\section{Learned Encoding Advisor}
\label{sec:approach}
LEA works in two phases, a training phase and an inference phase. In the training phase, synthetic data is used to pre-train LEA's internal models. LEA trains three models for each encoding type: one model to predict the encoded size, and two models which work together to predict the overall scan time for the column. The first model, which is the most expensive to train, can be trained offline on any hardware. The other two models capture details about the underlying hardware, and are thus trained \emph{in situ} on the target machine. 
In the inference phase, samples and statistics from the user's data are fed into those trained models.

Given a set of encodings and a specific objective, our goal is to find the optimal encoding for every column and block.
LEA allows for any objective function. Currently, it optimizes for either encoded size or from-storage scan speed (including in-memory decompression). In future work, we plan to investigate other potential objectives, include a mix between size and latency, and consider the impact of compression speed.

\sparagraph{Training Procedure.}
We generate slices (a column in a data block) of synthetic training data (more below), and extract from each slice hand-crafted features that are (1) easy to compute and (2) that we believe are highly relevant to encoding selection.
For integral data types, we collect the range, cardinality, and the first three moments (mean, variance, and skewness) of the distance between adjacent values within the slice.
For string slices, we collect the cardinality and mean string length.
Three models are trained for each encoding that predict the encoded size, in-memory scan speed, and from-storage scan speed of a slice. In addition, we obtain the encoded size of a contiguous 1\% sample for each encoding. The sample encoded size along with the shared slice statistics are used as model inputs.

Initially, we tried training LEA on several real datasets.
However, we found that doing so fails to provide a sufficient variety of training inputs and thus does not generalize well to unseen data.
Training LEA on a large number of real datasets (at least a few hundred columns in total) could mitigate that problem. 
However, we would still need to transfer all of these datasets onto the target system to obtain our measurements for the scan speed of each encoding.

Instead, we train LEA on data generated from synthetic distributions. 
For integers, we use (1) skewed normal distribution with randomly chosen mean and skewness, (2) discrete uniform distribution with randomly chosen cardinality, and (3) runs of the same value with randomly chosen run length.
For strings, we choose the mean length and cardinality of a slice and generate data based on those parameters.
A post-processing step randomly scales the data for integers, performs sorting, and inserts null values.
Cardinality, run length, and mean string length are selected from the log uniform distribution to more effectively explore the input space. 
Per data type (integral or string), LEA only needs to train on \textasciitilde 1000 slices each with 1\,M values for size, \textasciitilde 250 slices for in-memory scan speed, and \textasciitilde 5 slices for from-storage scan speed.

\sparagraph{Inference Procedure.}
Figure~\ref{fig:diagram} shows how LEA predicts encoded size, in-memory scan speed, and from-storage scan speed for a given encoding.
For the encoded size and in-memory scan speed models, we use random forest regression; due to various corner cases (\eg columns with very low cardinality), other regression techniques are not a good choice for the amount of training data we provide to the system. For strings longer than any in our training data, we use linear regression since random forests cannot extrapolate. From-storage scan speed uses linear regression to model the latency and throughput of the underlying storage device.

The entire inference process works as follows:
For each slice, we obtain the shared slice statistics.
Then, for each supported encoding, we take a 1\% contiguous sample of that slice (like during training) and encode it to measure the sample encoded size.
We feed both the slice statistics and the sample encoded size into the corresponding models for the encoding.
Finally, after predicting the properties of all encodings for a slice, we choose the encoding that performs best according to the given objective (either size or latency).

\section{Evaluation}
\label{sec:evaluation}

We evaluate LEA on top of a commercial column store (System C). Experiments are performed on an 8-core \verb|5d.2xlarge| AWS machine with a network-attached General Purpose SSD (\verb|gp2|) EBS device.
All from-storage experiments are with cold cache, \ie we clear the database and file system caches before each query.
We report the minimum latency from five runs.

\sparagraph{Encoding Strategies.}
System C offers 13 different column encodings.
All slices of a column are encoded using the same encoding.
System C comes with two built-in approaches for automatic encoding selection. The first approach, C-Default, uses a fixed mapping between encoding and data type (\eg string types always use run-length encoding). The second approach, C-Heuristic, is a heuristic-based encoding advisor that optimizes for encoded size (storage footprint).
C-Heuristic assumes that a lower storage footprint leads to better query performance.
We compare these two approaches with two variants of LEA: one that optimizes for encoded size (LEA-S) and one that optimizes for query latency (LEA-Q).

\sparagraph{Workloads.}
We evaluate these approaches using two different datasets.
StackOverflow is a denormalized table consisting of 12.5\,M rows and 24 columns.
We generate four queries that together cover most columns and perform various operations including selection, aggregation, and sorting.
Specifically, Q1 aggregates on a string column, Q2 aggregates and sorts on an intermediate numeric column, Q3 is an ungrouped aggregation, and Q4 is a string search in multiple columns.
The second workload operates on the full TPC-H dataset with scale factor 10.
We run all 22 TPC-H queries generated using \verb|QGEN| with default substitution parameters.

\begin{figure}
\includegraphics[width=1.0\columnwidth]{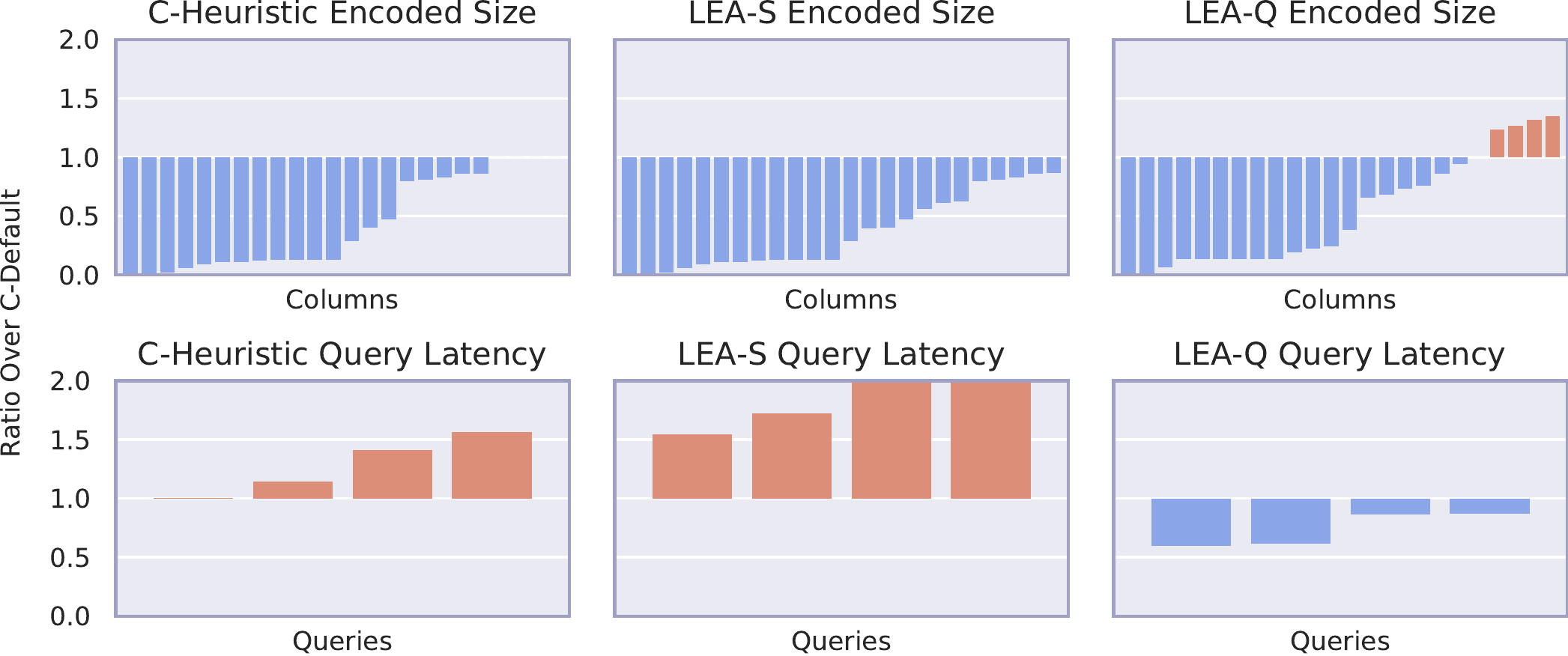}
\centering
\caption{Encoded size and query latency for StackOverflow.}
\label{fig:so-plot}
\end{figure}

\begin{figure}
\includegraphics[width=1.0\columnwidth]{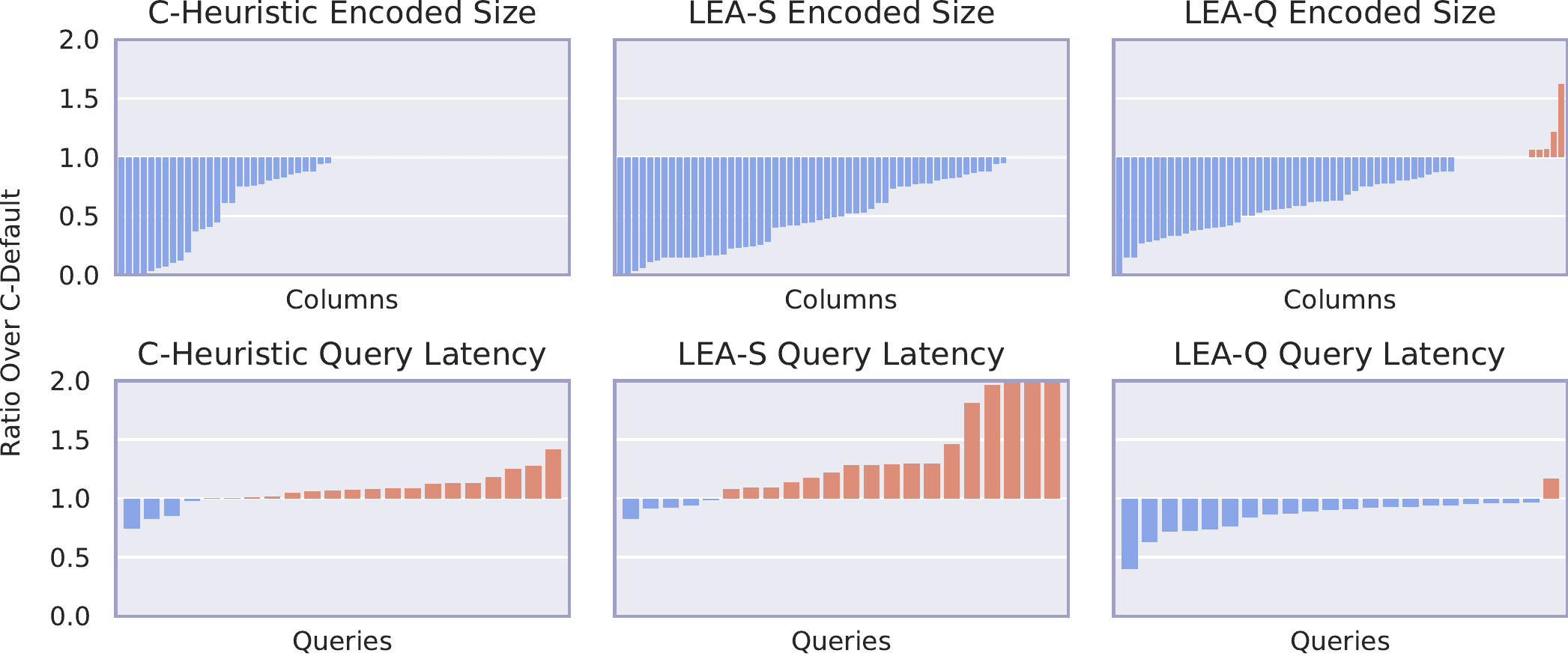}
\centering
\caption{Encoded size and query latency for TPC-H.}
\label{fig:tpch-plot}
\end{figure}

\sparagraph{Encoded Size and Query Latency.}
We integrate LEA with System C and report encoded size and cold-cache query latency.
Unfortunately, System C caches uncompressed blocks in memory, making warm cache experiments too complex for this work. 
Figure~\ref{fig:so-plot} shows the results on the StackOverflow workload.
Each of the three columns in the plot (C-Heuristic, LEA-S, and LEA-Q) shows the improvement/degradation ratio over System C's default encodings (C-Default).
The first row of plots shows encoded size and the second row shows query latency.
In terms of encoded size, all three approaches improve upon C-Default. Both C-Heuristic and LEA-S outperform C-Default on the 20 integral columns. However, only LEA-S produces smaller encodings on the four remaining string columns, while C-Heuristic does not.
In terms of query latency, LEA-Q improves upon the default encodings on all four queries, while LEA-S and C-Heuristic both show significant regressions.

Hence, we conclude that the size objective of LEA-S and C-Heuristic is not necessarily a good proxy for improved query performance.
While smaller columns are generally faster to load from storage, these time savings can be outweighed by the high CPU overhead that some of these encodings entail. 
Of course, this is hardware dependent; loading lightly compressed data from our network-attached disk (which has a maximum throughput of 250\,MiB/s) is faster than loading heavily compressed data and paying the higher CPU cost for decompression.
With even slower cloud storage, this trade-off might change. Traditional heuristics thus require re-tuning for each hardware configuration. Unlike traditional heuristics, LEA can adapt itself to \emph{the underlying hardware, the user's data, and the user's objective, all at the same time}~\cite{DBLP:conf/pldi/GottschlichSTCR18}.

Figure~\ref{fig:tpch-plot} shows the results for TPC-H.
C-Heuristic improves upon the default encodings in terms of encoded size for around half of the columns, while LEA-S improves encoded size for almost all columns.
LEA-Q again shows a few regressions but also shows the strongest improvements in terms of query latency.
In fact, it only slightly regresses on one of the 22 TPC-H queries while improving on the remaining queries by up to 50\%.
The regression on TPC-H Query 21 can be explained as follows: 
A sequential scan is performed on the same table three times, but only the first scan is from cold-cache.
System C is not able to cache all of the uncompressed blocks, so it has to effectively perform two additional in-memory scans over the encoded blocks.
LEA could estimate the cost of this query, but it would need to be provided with the column access patterns of this specific query or workload when selecting encodings.
Overall, LEA-Q achieves 19\% lower query latency while using 26\% less space than System C's heuristic-based encoding selection for this workload.
Similar to the StackOverflow results, solely optimizing for size (C-Heuristic and LEA-S) leads to query latency regressions on most queries and only improves latency on a few queries.

\begin{figure}
\centering
\includegraphics[width=0.8\columnwidth]{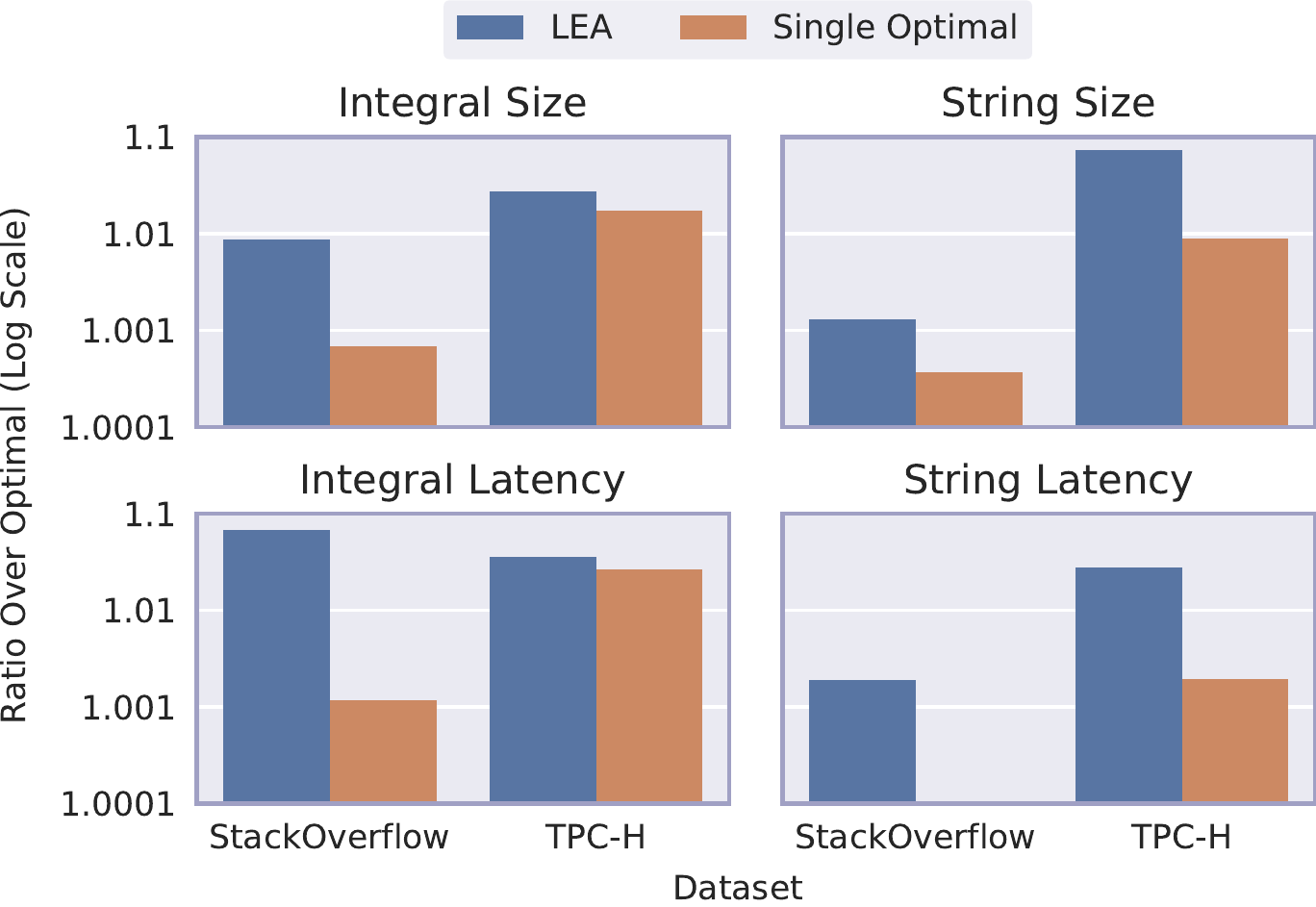}
\caption{Comparing LEA to the optimal set of column encodings (per slice). We also compare to the Single Optimal encoding (same encoding for all slices of a column).}
\label{fig:optimality}
\end{figure}

\sparagraph{Comparing Against Ground Truth.}
We now compare LEA's encoding recommendations to the ground truth, \ie the best possible set of column encodings for a given objective (size or query latency).
In this experiment, we perform full column scans instead of running specific queries to get a more complete picture.
So far, we were only able to choose one encoding per column (across all blocks).
However, real data may have localized correlations.
Hence, we now add an additional degree of freedom to allow for choosing an encoding per column \emph{and} block (\ie per slice).
Since System C does not support slice-specific encodings, we use our own prototype database for this experiment.
Our prototype supports delta, dictionary, frame-of-reference (FOR), run-length encoding (RLE), and ZSTD.
Since our engine currently does not support joins, we use a denormalized version of TPC-H at scale factor 1 with around 7\,M rows.
We brute force the optimal set of encodings (Optimal) for all slices (given the corresponding objective, size or query latency).
We compare Optimal with LEA as well as one more baseline: Single Optimal which brute forces the single best encoding for an entire column (and is hence not slice-specific).

Figure~\ref{fig:optimality} shows the results.
Across all configurations (datasets, data types, and objectives), LEA and Single Optimal are always within 10\% of Optimal.
Using one encoding per column is not the best in all cases because different blocks might contain different data distributions (\eg more duplicates).
While such patterns can be found in organic data, this effect is amplified by our use of \emph{denormalized} StackOverflow and TPC-H (\ie popular foreign keys are duplicated many times).
Although LEA is not able to outperform Single Optimal on these datasets, there is still value in being able to choose one encoding per slice, as evidenced by the difference between Single Optimal and Optimal.

\begin{figure}
\centering
\includegraphics[width=0.8\columnwidth]{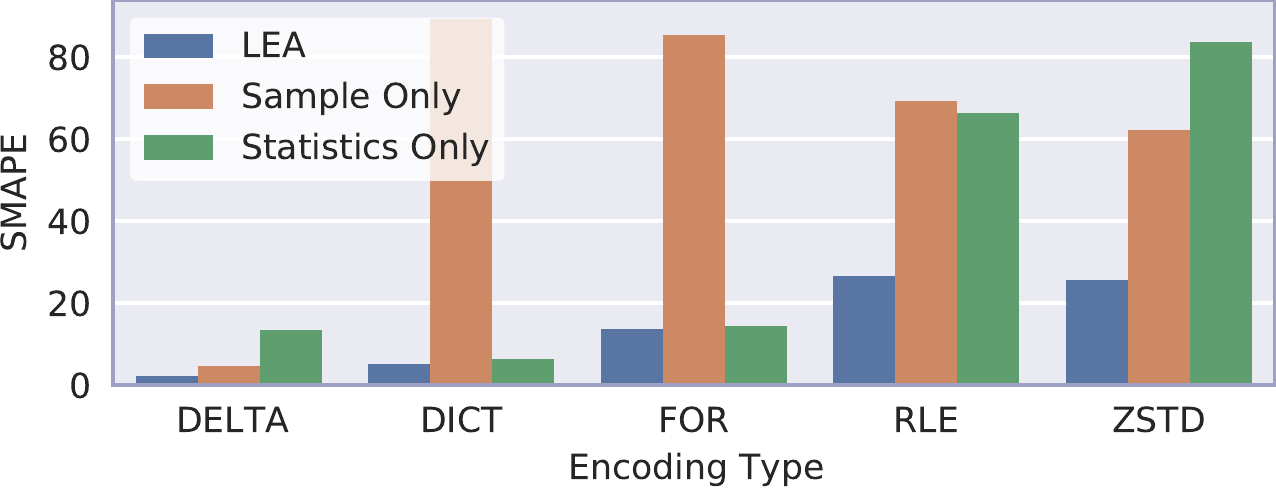}
\caption{Prediction accuracy of LEA-S compared to two ablated versions on denormalized StackOverflow and TPC-H.}
\label{fig:ablation}
\end{figure}

\sparagraph{Ablations.}
We now study the importance of individual model features on prediction accuracy.
We focus on integral columns and encoded size as the objective in this experiment.
We compare LEA (with sample and slice statistics) with two baselines: Sample Only which is LEA without slice statistics (such as min/max and cardinality) and Statistics Only which is LEA without sampling.
We use symmetric mean absolute percentage error (SMAPE) to compare predicted with actual sizes.
As shown in Figure~\ref{fig:ablation}, LEA outperforms both ablations across all encodings.
Slice statistics are particularly important for dictionary and FOR encodings. Dictionary encoding depends on cardinality and FOR encoding depends on domain size; however, both metrics are hard to estimate from a sample.

\begin{figure}
\centering
\includegraphics[width=0.8\columnwidth]{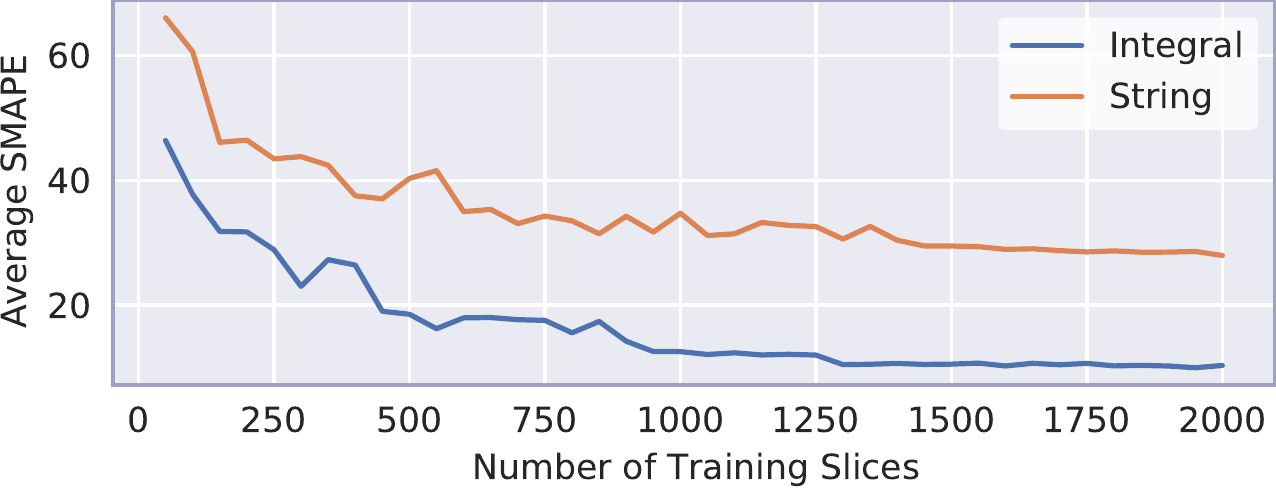}
\caption{Prediction accuracy of LEA-S with an increasing number of training slices on denormalized StackOverflow and TPC-H. SMAPE is averaged over supported encodings.}
\label{fig:mae}
\end{figure}

\sparagraph{Training Convergence and Time.}
Figure~\ref{fig:mae} shows the prediction accuracy of LEA-S with an increasing number of training slices.
Recall that each slice contains 1\,M rows, of which we take a contiguous 1\% sample.
We again use SMAPE to measure the difference between the predicted and actual sizes.
For both integral and string types, LEA requires around 1,500 slices to converge.
Notably, most training is spent obtaining the labeled training data.
With System C, our approach requires around three hours to load the various synthetic datasets with the different encodings and to collect the relevant metrics (encoded size and scan speed).
Training the random forest models takes less than one minute.

\sparagraph{Inference Time.}
Using LEA involves drawing slice and sample statistics from the data and feeding them to our different models. The most expensive step of the inference process is applying the different encodings to the sample.
For TPC-H, LEA on System C takes ten minutes for the entire encoding selection process, including the time it takes to load the data from all tables, gather statistics, and evaluate the models.
We expect that an optimized implementation integrated with System C would be able to reduce the inference time by an order of magnitude or more.
In contrast, brute-forcing the optimal set of encodings takes upwards of 30 minutes per column when scan speed needs to be measured.

\section{Conclusions}
\label{sec:conclusions}

We have presented a learned approach to encoding selection (LEA).
LEA is a first step towards an instance-optimized data encoding system that considers the given data, query workload, and hardware.
Trained with synthetic data and without workload knowledge, we have shown that LEA outperforms the auto-encoding heuristic implemented in a state-of-the-art commercial column store.
In future work, we plan to incorporate the concrete access patterns of a given query workload into the optimization process.

{
\setlength{\parskip}{0.5em}
\footnotesize
\sparagraph{Acknowledgments.}
This research is supported by Google, Intel, and Microsoft as part of DSAIL at MIT, and NSF IIS 1900933. This research was also sponsored by the United States Air Force Research Laboratory and the United States Air Force Artificial Intelligence Accelerator and was accomplished under Cooperative Agreement Number FA8750-19-2-1000. The views and conclusions contained in this document are those of the authors and should not be interpreted as representing the official policies, either expressed or implied, of the United States Air Force or the U.S. Government. The U.S. Government is authorized to reproduce and distribute reprints for Government purposes notwithstanding any copyright notation herein.
\par
}

\bibliographystyle{ACM-Reference-Format}

\end{document}